\documentclass{PoS}
\usepackage{tikz-feynman}

\title{$B_c \to B_{s(d)}$ form factors}

\ShortTitle{$B_c \to B_{s(d)}$}

\author{\speaker{Laurence Cooper}, Matthew Wingate\\
        Department of Applied Mathematics and Theoretical Physics, University of Cambridge, Cambridge, CB3 0WA, UK\\
        E-mail: \email{L.J.Cooper@damtp.cam.ac.uk, M.Wingate@damtp.cam.ac.uk}}

\author{Christine Davies, Judd Harrison\\
        SUPA, School of Physics and Astronomy, University of Glasgow, Glasgow G12 8QQ, UK\\
        E-mail: \email{Christine.Davies@glasgow.ac.uk, Judd.Harrison@glasgow.ac.uk}}
      
\author{Javad Komijani\\
	    SUPA, School of Physics and Astronomy, University of Glasgow, Glasgow G12 8QQ, UK\\
        Department of Physics, University of Tehran, Tehran 1439955961, Iran\\
        E-mail: \email{javad.komijani@glasgow.ac.uk}}
                
\author{HPQCD Collaboration}

\abstract{We present results of the first lattice QCD calculations of $B_c \to B_s$ and $B_c \to B_d$ weak matrix elements.
Results are derived from correlation functions computed on MILC Collaboration gauge configurations with lattice spacings between $0.12$ [fm] and $0.06$ [fm] including 2+1+1 flavours of dynamical sea quarks in the Highly Improved Staggered Quark (HISQ) formalism.
Form factors across the entire physical $q^2$ range are then extracted and extrapolated to the physical-continuum limit.
Two different formalisms are employed for the bottom quark: non-relativistic QCD (NRQCD) and heavy-HISQ.
Checking agreement between these two approaches is an important test of our strategies for heavy quarks on the lattice.}

\FullConference{37th International Symposium on Lattice Field Theory - Lattice2019\\
		16-22 June 2019\\
		Wuhan, China}

\graphicspath{{./}}
\begin{document}

\section{Introduction}

The semileptonic weak decays $B_c^+ \to B_s^0 \overline{l} \nu_l$ and $B_c^+ \to B^0 \overline{l} \nu_l$ proceed via tree-level flavour changing processes $c \to sW^+$ and $c \to dW^+$ parametrised by the Cabbibo-Kobayashi-Maskawa (CKM) matrix of the Standard Model.
Associated weak matrix elements can be expressed in terms of form factors which capture the non-perturbative QCD physics.
Precise determination of the normalisation and $q^2$ dependence of these form factors from lattice QCD will allow comparison with future experiment to deduce the CKM parameters $V_{cs}$ and $V_{cd}$.

The $B_c^+ \to B_s^0 \overline{l} \nu_l$ and $B_c^+ \to B^0 \overline{l} \nu_l$ decays involve the practical complication of a heavy spectator quark. 
Care must be taken in placing such a particle on the lattice to avoid large discretisation effects. 
We carry out one study with a valence NRQCD \cite{Lepage:1992tx} $b$ quark, allowing for computations with physically massive b quarks, and a complementary calculation using the fully relativistic approach of HPQCD's heavy-HISQ method \cite{McNeile:2010ji} which involves calculations for a set of quark masses on ensembles of fine lattices at a variety of lattice spacings, enabling a fit from which the physical result at the $b$ quark mass in the continuum can be determined.
The consistency of the NRQCD and heavy-HISQ approaches is demonstrated by comparing the form factors extrapolated to the physical-continuum limit.

The form factors $f_0$ and $f_+$ parametrise the continuum weak matrix element
\begin{equation} \label{form factors}
\langle  B_{s(d)} (\mathbf{p_2})  | V^{\mu} | B_c (\mathbf{p_1}) \rangle =  f_0 (q^2) \Bigg[ \frac{M_{B_c}^2 - M_{B_{s(d)}}^2}{q^2}q^{\mu} \Bigg] \nonumber \\
+ f_+(q^2) \Bigg[ p_2^{\mu} + p_1^{\mu} - \frac{M_{B_c}^2 - M_{B_{s(d)}}^2}{q^2}q^{\mu} \Bigg] 
\end{equation}
and are constructed from the matrix elements by fitting the correlator data to a sum of real exponentials, where $q = p_1 - p_2$ is the 4-momentum transfer.
By calculating correlators at a range of transfer momenta on lattices with different spacings and quark masses, continuum form factors at physical quark masses are obtained.

\section{Lattice Methodology}

\subsection{Lattice Parameters}
Ensembles with $2+1+1$ flavours of HISQ sea quark generated by the MILC collaboration \cite{Bazavov:2010ru,Bazavov:2012xda,Bazavov:2015yea} are described in table \ref{LattDesc1}.
The Symanzik improved gluon action used is that in \cite{Hart:2008sq} where the gluon action is improved perturbatively through $\mathcal{O}(\alpha_s a^2)$ to account for dynamical HISQ sea quarks.
HISQ \cite{Follana:2006rc} is used for all other valence flavours.
Masses used for the HISQ propagators calculated with the MILC code \cite{MILCgithub} on these gluon configurations are tabulated here also.
Our calculations feature physically massive strange quarks and equal mass up and down quarks, with a mass denoted by $m_l$, with $m_l/m_s = 0.2$ and also the physical value $m_l/m_s = 1/27.4$ \cite{Bazavov:2014wgs}.

\begin{table}
\begin{center}
 \begin{tabular}{c c c c c c c} 
 \hline\hline
set & $w_0/a$ & $N_x^3 \times N_t$ & $n_{cfg}$ & $am_l^{{sea}}$ & $am_s^{{sea}}$ & $am_c^{{sea}}$ \\ [0.1ex] 
\hline
1 & 1.1119(10) & $16^3 \times 48$ & $1000$ & $0.013$ & $0.065$ & $0.838$ \\
2 & 1.1367(5) & $32^3 \times 48$ & $500$ & $0.00235$ & $0.00647$ & $0.831$ \\
3 & 1.3826(11) & $24^3 \times 64$ & $1053$ & $0.0102$ & $0.0509$ & $0.635$ \\
4 & 1.4149(6) & $48^3 \times 64$ & $1000$ & $0.00184$ & $0.0507$ & $0.628$ \\
5 & 1.9006(20) & $32^3 \times 96$ & $504$ & $0.0074$ & $0.037$ & $0.440$ \\
6 & 2.896(6) & $48^3 \times 144$ & $250$ & $0.0048$ & $0.024$ & $0.286$ \\
 \hline\hline
\end{tabular}
\caption{Parameters for the MILC ensembles of gluon field configurations. The lattice spacing $a$ is determined for the Wilson flow parameter $w_0$ given in lattice units for each set in column 2 where values were obtained from \cite{Chakraborty:2016mwy} on sets 1 to 5 and \cite{Chakraborty:2014aca} on set 6. The physical value $w_0 = 0.1715(9)$ was fixed from $f_{\pi}$ in \cite{Dowdall:2013rya}. The very-coarse lattices, sets 1 and 2, have $a \approx 0.15$ [fm], and the coarse lattices, sets 3 and 4, have $a \approx 0.12$ [fm]. Sets 5 and 6 have $a \approx 0.09$ [fm] and $a \approx 0.06$ [fm] respectively. Sets 1, 3, 5 and 6 have unphysically massive light quarks such that $m_l/m_s = 0.2$. 
Sets 1 to 5 were used in the NRQCD calculation of the form factors. The heavy-HISQ calculation used sets 3, 5 and 6. 
}

\label{LattDesc1}
\end{center}
\end{table}

We work in the frame where the $B_c^+$ is at rest, and momentum is inserted into the strange and down valence quarks through twisted boundary conditions \cite{Sachrajda:2004mi} in the $(1\hspace{1mm}1\hspace{1mm}1)$ direction.
For the heavy-HISQ calculation, we use heavy quark masses up to $am_h = 0.8$.


\subsection{Correlators} \label{2ptcorrels}
\begin{figure} 
\begin{center}
\begin{tikzpicture}[scale=1.0, transform shape]
\begin{feynman}
	\vertex (A1);
	\vertex[below=2.5em of A1] (A2);
	\vertex[left=0.1em of A2] (src){\( B_c\)};
        \vertex[right=0.2em of A2] (src2);
	\vertex[below=4em of A2] (A3);
	\vertex[right=10em of A1] (B1);
	\vertex[below=4em of B1] (B2);
	\vertex[below=3em of B2] (B3);
         \vertex[below=0.2em of B1] (curr);
	\vertex[right=10em of B1] (C1);
	\vertex[below=2.5em of C1] (C2);
        \vertex[right=0.1em of C2] (snk){\( B_{s(d)} \)};
        \vertex[left=0.2em of C2] (snk2);
	\vertex[below=5em of C2] (C3);
	\vertex[above=1.5em of A1] (A0){0};
	\vertex[above=1.5em of B1] (B0){\( t\)};
	\vertex[above=1.5em of C1] (C0){\( T\)};
    
     \diagram* {
      {[edges=fermion]
        (C2) -- [with arrow=0.5, out=-150, in=-30, edge label'=\(\overline{b}\)](A2),
        (A2) -- [with arrow=0.5, out=30, in=180, insertion=0.9999, edge label'=\(c\)](B1) [crossed] -- [with arrow=0.5, out=0, in=150, edge label'=\(s (d)\), momentum={[arrow shorten=0.35] \( q\)}](C2),
      },
      (C0) -- [scalar] (C2) 
      (A0) -- [scalar] (A2),
      (B0) -- [scalar] (B1),
    };
  \end{feynman}
\end{tikzpicture}
\caption{3-point correlator $C_{{3pt}} (t,T)$. The operator insertion is denoted by a cross at timeslice $t$.}
\label{3ptcorrelfig}
\end{center}
\end{figure}
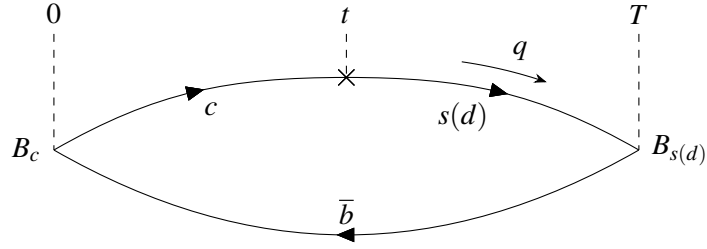

Random wall source \cite{Aubin:2004fs} HISQ propagators are combined with random wall source NRQCD $b$ propagators to generate $B_c^+$ and $B_{s(d)}^0$ 2-point correlator data. 
The HISQ charm propagator in the 3-point correlator, represented diagrammatically in figure \ref{3ptcorrelfig}, uses the random wall bottom propagator as a sequential source.

The correlators are fit to the following functions through use of the \textit{corrfitter} package \cite{corrfitter}.
The fit seeks to minimise an augmented $\chi^2$ as described in \cite{Lepage:2001ym,Hornbostel:2011hu,Bouchard:2014ypa}.
The functional forms 
\begin{eqnarray} \label{corrfitform}
C^{B_{s(d)}}_{2pt} (t) &=& \sum_i a[i]^2 e^{-E_a[i] t} - \sum_i a_o[i]^2 (-1)^t e^{-E_{a_o}[i]t} \nonumber \\
C^{B_c}_{2pt} (t) &=& \sum_j b[j]^2 e^{-E_b[j] t} - \sum_j b_o[j]^2 (-1)^t e^{-E_{b_o}[j]t} 
\end{eqnarray}
\begin{eqnarray}
C_{3pt} (t,T) = \sum_{i,j} a[i] e^{-E_a[i]t}  V_{nn}[i,j] b[j] e^{-E_b[j](T-t)} - \sum_{i,j} (-1)^{T-t} a[i] e^{-E_a[i]t}  V_{no}[i,j] b_o[j] e^{-E_{b_o}[j](T-t)} \nonumber \\
\hspace{0mm}- \sum_{i,j} (-1)^t a_o[i] e^{-E_{a_o}[i]t}  V_{on}[i,j] b[j] e^{-E_b[j](T-t)} + \sum_{i,j} (-1)^T a_{o}[i] e^{-E_{a_o}[i]t}  V_{oo}[i,j] b_o[j] e^{-E_{b_o}[j](T-t)} \nonumber \\
\end{eqnarray}
follow from the spectral decomposition of the Euclidean correlators with additional oscillatory contributions due to the coupling of different tastes of staggered quark.
The matrix elements are related to the fit parameters $V_{nn}[0,0]$ through $\sqrt{2E_{B_{s(d)}} 2E_{B_c}} V_{nn} [0,0] = \langle B_{s(d)} | J | B_c \rangle$, where $J$ is the relevant operator that facilitates the $c \to s(d)$ flavour transition.
On each set, the 2-point and 3-point correlator data for both $c \to s$ and $c \to d$ at all momenta is fit simultaneously to account for all possible correlations. Matrix elements and energies are then extracted.


\subsection{Extracting the form factors}
For both HISQ and NRQCD spectator quarks, the HISQ action is used for the quarks that participate in the current.
Hence, the current normalisation can be determined non-perturbatively by making use of the Partially Conserved Vector Current (PCVC) Ward identity
\begin{equation}\ \label{contPCVC}
\partial_{\mu} V^{\mu} = (m_c - m_{s(d)}) S
\end{equation}
relating the conserved (point-split) $c \to s(d)$ lattice vector current and the local lattice scalar density $S$. We choose local lattice operators only, thus equation (\ref{contPCVC}) must be adjusted by a single renormalisation factor $Z_V$ associated with the local lattice vector current giving
\begin{equation}
q_{\mu} \langle B_{s(d)} | V^{\mu} | B_c \rangle Z_V = (m_c - m_{s(d)})  \langle B_{s(d)} | S | B_c \rangle.\label{PCVClatt}
\end{equation}
Combining equations (\ref{form factors}) and (\ref{PCVClatt}) gives a determination of $f_0$  solely in terms of the scalar density matrix element through 
\begin{equation}
f_0 \big(q^2 \big) = \langle B_{s(d)} | S | B_c \rangle \frac{m_c - m_{s(d)}}{M_{B_c}^2 - M_{B_{s(d)}}^2}. \label{f0scalar}
\end{equation}
Thus, we are concerned with insertions of the local scalar density $J=S$ as well as the local vector current $J=V$.
Once $f_0$ is determined, $f_+$ is obtained using equation (\ref{form factors}).


\section{Results}

\begin{figure}
	\begin{center}
		\includegraphics[width=0.55\textwidth]{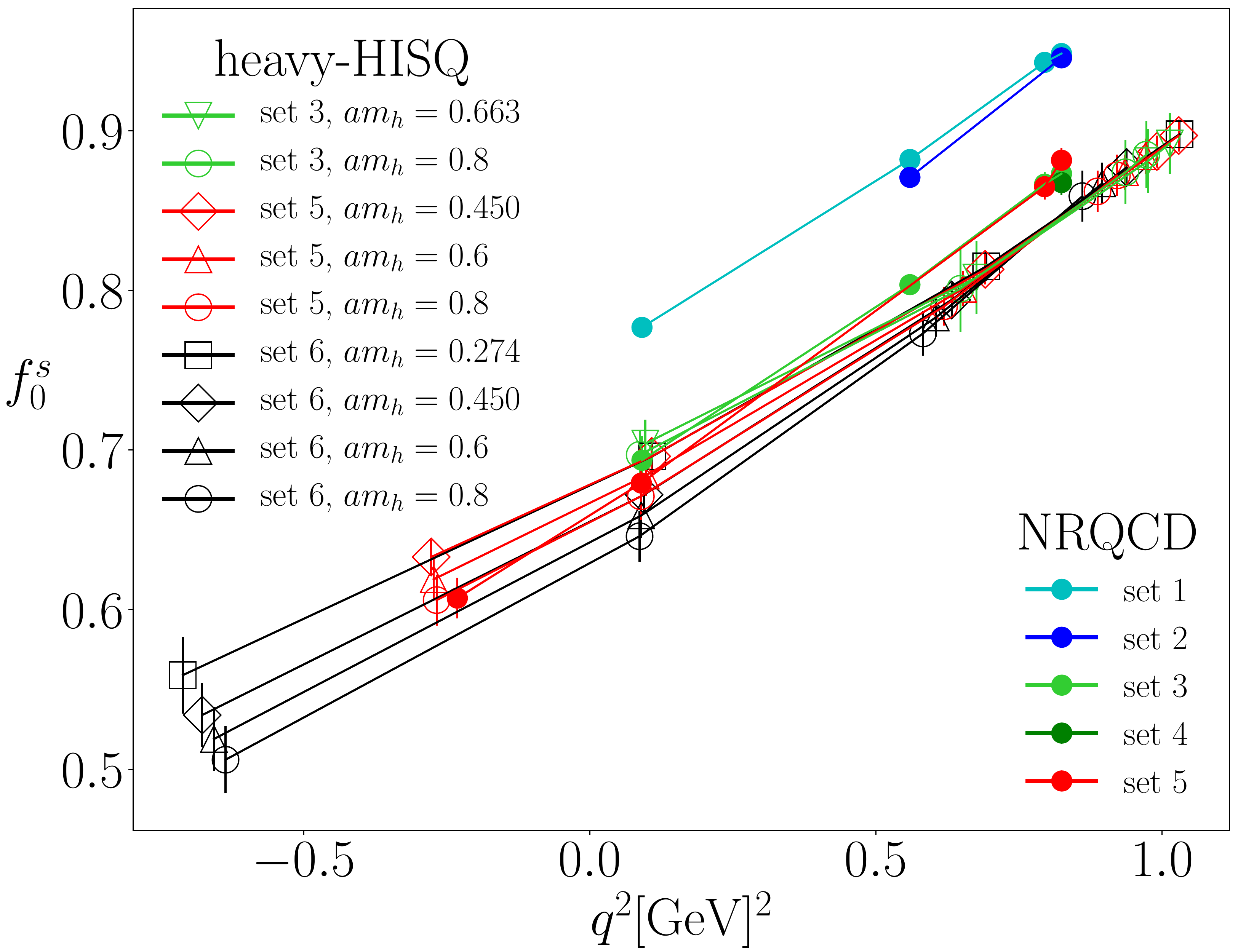}
		\caption{$f_0$ form factor data for $B_c^+ \to B_s^0 \overline{l} \nu_l$ from both the NRQCD and heavy-HISQ approaches. Filled in circles denote NRQCD form factor data.}
		\label{raw_f0_comp_ctos}
	\end{center}
\end{figure}

Figure \ref{raw_f0_comp_ctos} shows data for the form factor $f_0$ for the  $B_c \to B_s$ process.
The data for all momenta on all the lattices is fit simultaneously to a functional form which allows for dependence on the lattice spacing $a$ and mistuned bare quark masses.
The fit is carried out using the \textit{lsqfit} package \cite{lsqfit} that implements a least-squares fitting procedure.
It is convenient to map the semileptonic region $0<q^2<(M_{B_c} - M_{B_{s(d)}})^2$ to within the unit circle through
\begin{eqnarray} \label{littlez}
t_{\pm} &=& (M_{B_c} \pm M_{B_{s(d)}})^2, \nonumber \\
z &=& \frac{\sqrt{t_+ - q^2} - \sqrt{t_+ - t_0}}{\sqrt{t_+ - q^2} + \sqrt{t_+ - t_0}},
\end{eqnarray}
so that the form factors can be approximated by a truncated power series in $z$.
The parameter $t_0$ is chosen to be 0, thus the points $q^2 = 0$ and $z = 0$ coincide.
Expressing the form factor as a polynomial in $z$ was a suitable approach for $D \to K(\pi)$  in \cite{Koponen:2013tua} since the polynomial coefficients where $\mathcal{O}(1)$. 
The value $z(q^2 = t_-)$ is two orders of magnitude smaller than in \cite{Koponen:2013tua}, yet the ranges of physical $q^2$ are comparable since $t_-^{B_c \to B_{s(d)}} /  t_-^{D \to K} = \mathcal{O}(1)$.
Hence, $z$ in equation (\ref{littlez}) must be rescaled appropriately to ensure that the polynomial coefficients are $\mathcal{O}(1)$, a desirable property when setting prior distributions.
In this study, we rescale $z$ and define $z_p (q^2) = z(q^2)/|z(M_{res}^2)|$, where $M_{res}$ is the mass of the nearest resonance.

With an NRQCD spectator quark, the form factor fit takes the form
\begin{eqnarray} \label{fffitform}
f(q^2) = & \hspace{1mm} P(q^2) \sum\limits_{n=0}^{3} b^{(n)} z_p^n,
\end{eqnarray}
where the pole structure is represented by a factor $P(q^2) = (1 - q^2 / M_{res}^2)^{-1}$ multiplying a polynomial whose coefficients are
\begin{eqnarray} \label{fffitform2}
b^{(n)}= & \hspace{1mm} A^{(n)} \Big\{ 1+ B^{(n)} (am_c/\pi)^2 + C^{(n)} (am_c/\pi)^4 \Big\} 
\end{eqnarray}
with further terms that account for quark mass mistunings.

The heavy-HISQ data requires a fit form that accounts for $(am_h)^{2n}$ discretisation effects as well as physical dependence on $m_h$. Motivated by HQET we express this physical heavy mass dependence as a power series in $\Lambda_{QCD}/M_{H_c}$.
The form factor data from heavy-HISQ is fit to
\begin{eqnarray}
f (q^2) = P(q^2)\sum_{n,i,j,k = 0}^{3} A^{(n)}_{ijk} z_p^n \left(\frac{am_c}{\pi}\right)^{2i} \left(\frac{am_b}{\pi}\right)^{2j} \Delta_{H_c}^{(k)},
\end{eqnarray}
where, for $k=0$, $\Delta_{H_c}^{(k)}=1$ and, for $k \neq 0$,
\begin{equation}
\Delta_{H_c}^{(k)} =\left(\frac{\Lambda_{QCD}}{M_{H_c}}\right)^k- \left(\frac{\Lambda_{QCD}}{M_{B_c}}\right)^k .
\end{equation}
The mistuning terms are again omitted for brevity.
Finally, the kinematic relation $f_0 (0) = f_+ (0)$
is imposed on the fit as a constraint alongside the data.

\begin{figure}
\begin{center}
\includegraphics[width=0.49\textwidth]{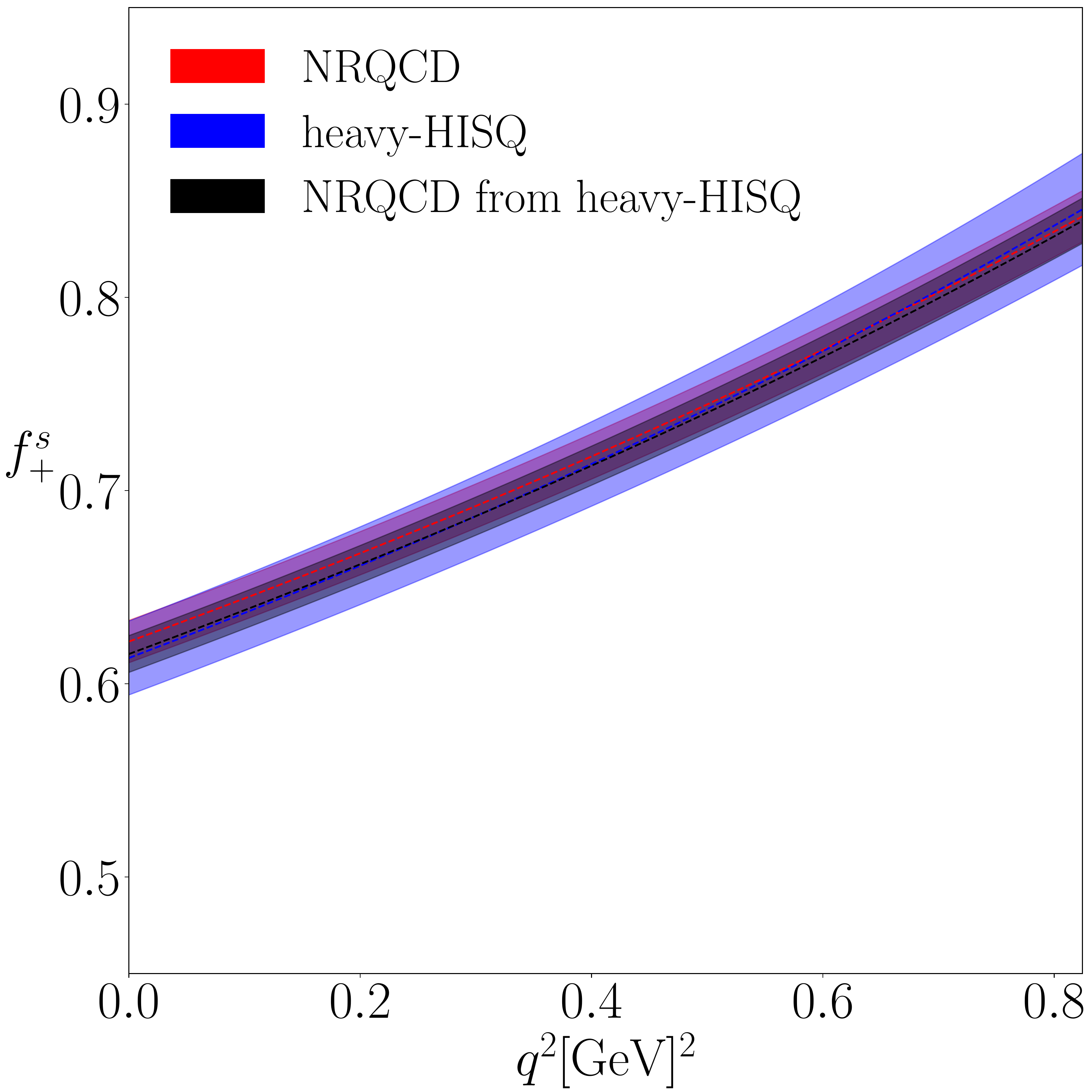}
\includegraphics[width=0.49\textwidth]{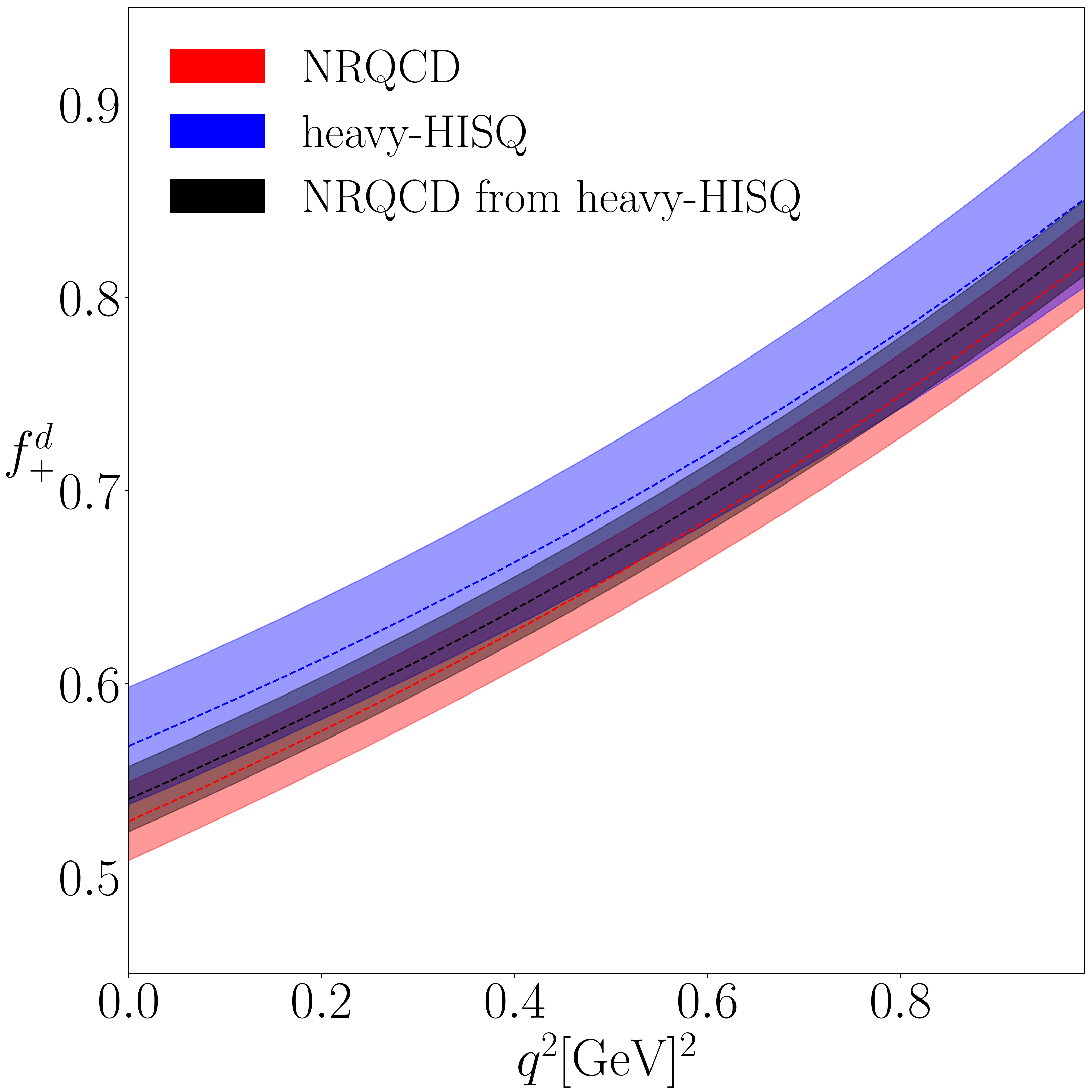}
\caption{Fits of $f_+$ for $B_c^+ \to B_s^0 \overline{l} \nu_l$ and $B_c^+ \to B^0 \overline{l} \nu_l$ tuned to the physical-continuum limit. The black band is the fit where results from the fit of the heavy-HISQ data are used as priors for the fit of form factor data with an NRQCD spectator quark.}
\label{fplus_physcont_comp_strange}
\end{center}
\end{figure}

The form factor $f_+$ for the $B_c \to B_s$  and $B_c \to B_d$ processes, at the physical-continuum limit, from NRQCD and heavy-HISQ is shown in figure \ref{fplus_physcont_comp_strange}.
Plotted alongside the functions from the heavy-HISQ and NRQCD calculations is a function arising from a chained fit where the $A_{000}^{(n)}$ from the heavy-HISQ fit were used as prior distributions for the $A^{(n)}$ in the form factor fit forms in the NRQCD study.
This chained fit has $\chi^2/$d.o.f. $= 1.4$ and is consistent with both the separate fits.
The chained fit is labelled `NRQCD from heavy-HISQ' in figure \ref{fplus_physcont_comp_strange}.

\section*{Acknowledgments}

We are grateful to Mika Vesterinen for asking us about the form factors for these decays at the UK Flavour 2017 workshop at the IPPP, Durham.
We are also grateful to Matthew Kenzie for discussions about the prospects of measurements by LHCb.
We thank Jonna Koponen, Andrew Lytle and Andre Zimermmane-Santos for making previously generated lattice propagators available for our use.
We thank the MILC collaboration for making publicly available their gauge configurations and their code MILC-7.7.11 \cite{MILCgithub}.
This work was performed using the Cambridge Service for Data Driven Discovery (CSD3), part of which is operated by the University of Cambridge Research Computing on behalf of the STFC DiRAC HPC Facility (www.dirac.ac.uk).
The DiRAC component of CSD3 was funded by BEIS capital funding via STFC capital grants ST/P002307/1 and ST/R002452/1 and STFC operations grant ST/R00689X/1.
DiRAC is part of the National e-Infrastructure.
We are grateful to the CSD3 support staff for assistance.
This work has been partially supported by STFC consolidated grant ST/P000681/1.

\bibliographystyle{JHEP}
\bibliography{Bc_Bsd}
%

\end{document}